# Near-infrared integral-field spectroscopy of HD209458b

Daniel Angerhausen[*a], Alfred Krabbe[a], Christof Iserlohe[a],

[a] I. Physikalisches Institut, University of Cologne, Zülpicher Str.77, 50937 Köln, Germany;

## ABSTRACT

We present first results of an exploratory study to use integral field spectroscopy to observe extrasolar planets. We focus on transiting 'Hot Jupiters' and emphasize the importance of observing strategy and exact timing. We demonstrate how integral field spectroscopy compares with other spectroscopic techniques currently applied. We have tested our concept with a time series observation of HD209458b obtained with SINFONI at the VLT during a superior conjunction.

## 1. INTRODUCTION

The search for exoplanets and their characteristics has become a fast-growing field in astrophysics. Most of the about 150 known exoplanets have been found indirectly (Marcy et al. 1995) via dynamical effects (Doppler-shift of stellar lines). Several so called transiting exoplanets have been detected due to variations in the stellar light curves (Henry et al. 2000, various surveys). These exoplanets have their orbits inclined so that they move in front of the stellar disk during inferior conjunction (IC) and disappear behind the host star during superior conjunction (SC). First direct detection of infrared light from an transiting extrasolar planet was recently demonstrated (Deming et al. 2005, Charbonneau et al.

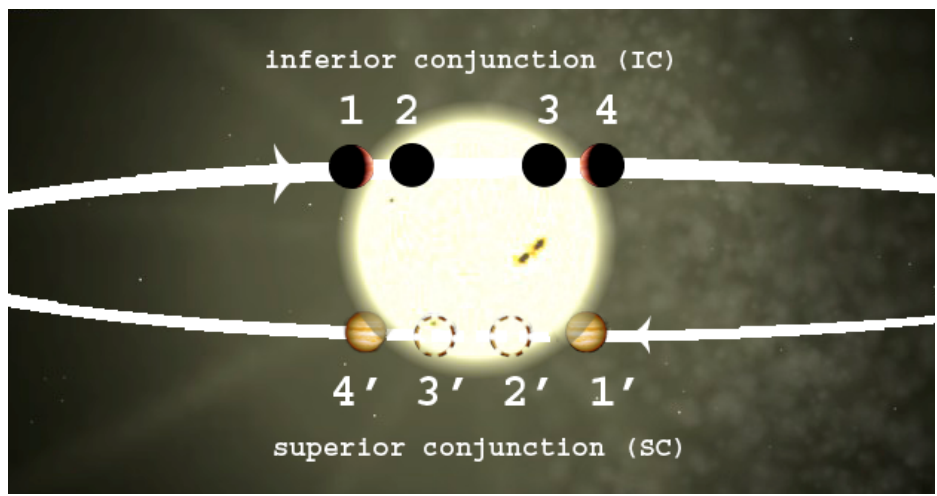

Fig 1. The secondary eclipse starts with the planet disappearing behind the host star. The midpoint between 2nd and 3rd contact is the superior conjunction (SC). If spectra of the (spatially unresolved) stellar light are obtained under stable atmospheric conditions, the difference between the stellar spectrum before disappearance or after recurrence, and the spectrum while the planet orbits behind the star, should reveal the planetary spectrum plus stellar light reflected by the planetary atmosphere.

[*] anger@ph1.uni-koeln.de

2005). Although imaging of extrasolar planets already provides valuable information about their spectral energy distribution (SED), the important step is obtaining spectra of exoplanets to investigate their composition and atmospheres.

Already a simple model for computing contrast and expected signal to noise ratio (S/N) shows that the subgroup of transiting exoplanets has the lowest star/planet contrast ratio and some members should already be observable with near infrared (NIR) integral field spectroscopy (IFS) from large ground based telescopes. The most promising of these candidates so far identified is the transiting 'Hot Jupiter' exoplanet HD 209458b. Its orbit has already been determined with sufficient accuracy and its orbital period of about 3.5 days provides frequently recurring conjunctions.

The technique of choice for spectroscopy of transiting extrasolar planets is to cancel the stellar contribution by using a differential method: Before the planet is disappearing or after its recurrence the observable signal is the sum of the light from the planet, the star, and various background contributions. During the superior conjunction, only the planetary contribution is missing. The subtraction between both observations should reveal the planetary spectrum plus stellar light reflected off the planetary atmosphere (Fig. 1). The reflected light will be accounted for when the difference is divided by the stellar spectrum for the purpose of spectral calibration. The crux of this observing technique is that one needs to build up sufficient S/N of the subtrahend during the time of superior conjunction. In contrast, the S/N of the minuend is generally not critical since before and after occultation plenty of time usually is available. While the timing issue will be addressed in paragraph 3.2, it should already be noted here that the duration of occultation for HD 209458b is slightly more than 2 hours, which seems just feasible.

In section 2 we address the specific advantages of integral field spectroscopy for this type of observations. Section explains the observing strategy, while section 4 focuses on the data reduction. Section 5 summarizes our conclusions.

## 2. INTEGRAL FIELD SPECTROSCOPY

Integral field spectroscopy (IFS) is a method to simultaneously measure the spectra at many locations arranged in a rectangular or other 2-d pattern on the sky. Because of the 3-d format of the information to be obtained with a 2-d detector device, integral field spectrographs generally have significantly smaller fields of view than pure imagers and are ideally suited for relatively small single objects like host stars plus exoplanet(s). IFS has several advantages over other techniques applied by, e.g., Richardson et al. (2003): Integral field spectrographs like SINFONI used here are by a factor of 2 - 3 more sensitive compared with single slit spectrographs, which is mostly due to lower slit losses. They also minimize systematic noise and efficiently operate in the NIR at the diffraction limit of modern 8-10m class telescopes. Also, both, the planetary spectrum as well as all related spatial information, are conserved within a consistent 3-dim data cube and can thus be analyzed together. Although the host stars of exoplanets are usually bright, the G0V star HD209458 has $m_V$ = 7.65 mag, high spectral resolution near infrared observations can still cope with the photon flux without saturating the detector within several tens of seconds.

The combination of IFS and adaptive optics (AO) is particularly desirable, although not intuitively obvious. However, several arguments favor such a combination: The smaller angular scale substantially lowers the time scale of spatial modulations of the background across the FOV. Variable spatial modulations in the background, which is one of the main contributors to spatial systematic noise across the field of view (FOV), is thus minimized. The PSF becomes much more stable and reproducible, thus lowering contributions from speckle noise. The smaller angular image scale also lowers the amount of thermal background on the detector pixel in the long part of the K band, thus improving the S/N. Finally, dithering steps for minimizing pattern noise are much better controllable using AO. In combination with AO in the H or K-band, SINFONI is in fact one of the most sensitive existing NIR instrument to detect faint targets showing characteristic intrinsic spectral features.

Cancellation of residual systematic speckle noise has recently been attempted with some success by Hartung et al. (2004) using differential spectral imaging with narrow band filters. Complete cancellation of the residual systematic noise, however, can only be achieved with an imaging spectrograph provided that the basic reduction steps have been thoroughly applied. It also helps if the intrinsic spectrum of HD209458b shows significant structure, which has been predicted by, e.g., Sudarsky, Burrows & Hubeny (2003) but which still need to be proven.

## 3. OBSERVATION

SINFONI is the combination of the integral field unit SPIFFI (SPectrograph for Infrared Faint Field Imaging) and the curvature sensing adaptive optics module MACAO (Multi-Application Curvature Adaptive Optics). The SINFONI (Spectrograph for Integral Field Observation in the Near-Infrared) field of view on the sky is sliced into 32 slitlets. Each of the 32 slitlets is reimaged onto a 64-pixel wide strip of the detector. Thus, SINFONI obtains 32 x 64 = 2048 spectra of the observed region of the sky. For our observation pre-optics were set up to a scale of 100 mas for the widths of the slices providing a field of view (FOV) of 3''x 3''. Typical limiting magnitudes (S/N=10 in 1h on source) are around 17-18 mag in J, H, K (Kissler-Patig 2005).

### 3.1 Observing strategy

The observation of HD209458b is constrained by several parameters. A superior conjunction (SC) of HD209458b can be observed about once a week due to the rotation period of 3.525 days. Due to the SC shift of 1.2 hours/week, a similar observing pattern reoccurs every 10 weeks. The maximum elevation of HD209458b at the VLT is about 47º, limiting the observation to the time around local transit at the observatory.

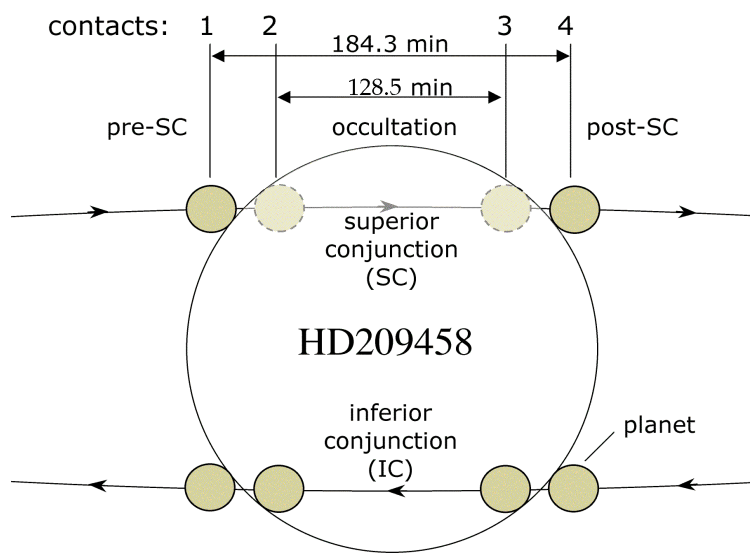

Fig. 2 The relative timing of the four contacts of the superior conjunction of HD 209458b.

Matching the S/N of both phases, during occultation between the $2^{nd}$ and the $3^{rd}$ contact and outside occultation before and after the $1^{st}$ and $4^{th}$ contact requires a minimum observing time on target of 317 min already exceeding 5 hours (Fig. 2). It is therefore highly desirable to observe the target around opposition, when it is observable for at least seven hours. Matching these three constraints results in very good observing conditions of about 2 nights/ year, each a week apart and maybe a third night under still acceptable conditions (e.g. lower mean elevation). A more general investigation reveals that the same is also valid for other large observatories, e.g., for Hawaii, where the target transits close to zenith.

The observation conditions can be optimized by analyzing the observing strategy one step further. Since the target will be raising or setting or both during the observations, the SC phase relative to the transit still is a free parameter. There is no intrinsic gain distributing the observing time symmetrically with respect to SC midpoint. Rather, the goal is to obtain two sets of data under mostly identical atmospheric conditions. Ideally, local culmination of HD209458b should occur between contacts 1 or 2 or between contacts 3 and 4 (as in Fig.4). Then the range of zenith distance covered by both observations is mostly identical, thus improving the data quality significantly. The optimal constellation for observing HD209458b then in fact occurs only once or twice per year, strongly limiting the observing opportunities.

## 3.2 Timing

Exact timing is critical since the observation time during the SC phase is limited. Model calculations show that under good atmospheric and instrumental conditions observations of both phases during a single SC event result in a spectrum with a total S/N of about 40. The exact times of the superior conjunctions were still uncertain at the time of the observations. We assessed them from the times of inferior conjunctions, which have been determined quite frequently, from the best determination of the orbital eccentricity and from the longitude of periastron (Wittenmyer et al. 2004,

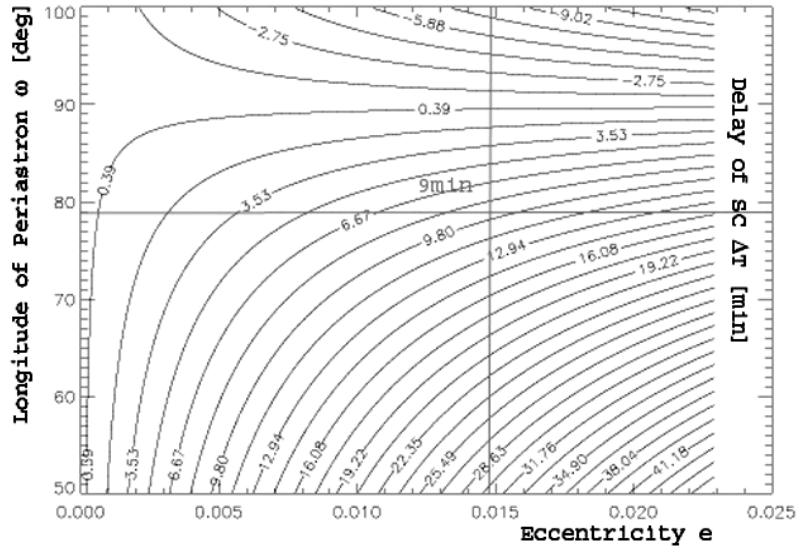

Fig. 3 Dependence of delay of SC on longitude of periastron and the orbital eccentricity of HD209458b based on the best data available in spring 2005. This diagram was used for the determination of the offset between the time of the inferior conjunctions plus half an orbital period and the superior conjunctions. We determine the offset to 9 ± 20 minutes, shown at the crossing of the two straight lines.

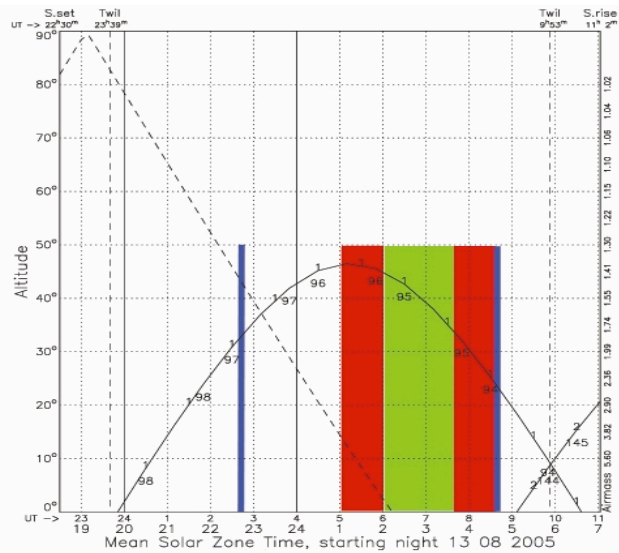

Fig.4. Time vs. Altitude/Airmass diagram of our SINFONI-VLT observation on Aug. 13 2005. the thick vertical bars represent start and end of our observation. The light shaded area marks the phase of secured conjunction (between 2$^{nd}$ and 3$^{rd}$ contact). The shaded areas adjacent represent phases of ingress (between 1$^{st}$ and 2$^{nd}$ contact) and egress (between 3$^{rd}$ and 4$^{th}$ contact) as well as uncertainties of delay and duration of the SC (Created with http://www.ing.iac.es/ds/staralt/index.php).

Laughlin et al. 2005). Fig. 3 shows the dependence of the delay of the SC on longitude of periastron and the eccentricity based on the best data available in spring 2005. The delay Δt of the SC is defined as $t_{SC} - t_{IC} - t_{orbit}/2$ where the selected IC is immediately preceding the SC. From the diagram we determined the offset to 9 ± 20 minutes, shown at the crossing of the two straight lines. Meanwhile, based on Deming et al (2005), who was the first to observe a secondary transit of HD209458b, the offset has been determined to 0 ± 7 min, very close to our estimation. The secondary eclipse pretty much occurs at the midpoint between transits of the planet in front of the star, which means that a dynamically significant orbital eccentricity is unlikely. The updated SC timing was included into our data reduction concept.

The optimal observing setting occurred for the ESO-Paranal observatory on August 13$^{th}$ 2005 as illustrated in Fig. 4. Our application for directors discretionary time was approved and we were awarded 7 hours observing time (as shown in Figure 4) on HD 209458b using the VLT-SINFONI instrument plus AO. Under sub-optimal atmospheric conditions we obtained 38 minutes worth of on target data during pre-SC phase and 60 minutes during SC phase as well as the corresponding sky frames for background correction and some telluric standard stars for atmospheric correction.

Clouds passed through several times during observation resulting in significant fluctuations of the atmospheric transmission (Fig. 5 and Fig. 8). Correction of atmospheric changes exhibited one of the biggest challenges during data reduction. SINFONI operated in H+K mode where a complete H+K spectra at a spectral resolution of R = 1500 are imaged on the detector. The target was dithered on the detector between two fixed positions in an ABBA fashion.

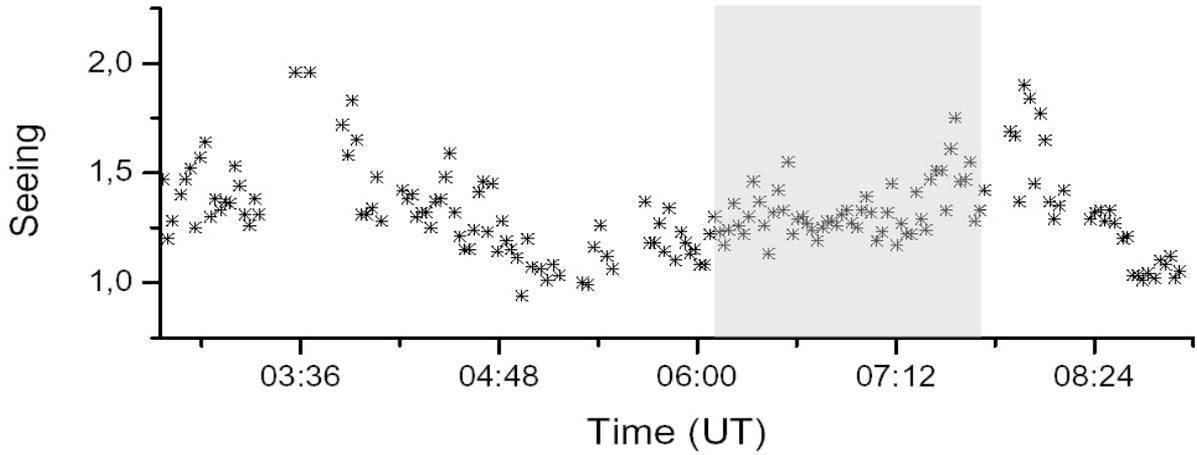

Fig.5. The visual seeing determined by the observatory varied between 1" and 2" during the observations.

## 4. DATA REDUCTION CONCEPT

### 4.1 Extraction of spectra from the data cubes

In order to collect as much signal as possible, sky frames were only obtained for each dithered pair of on-target frames. Therefore, for each data frame the appropriate sky frame was constructed by time stamp sensitive linear interpolation. After sky subtraction, the frames were divided by a flat-field, followed by a procedure to detect bad pixel and cosmics on each individual frame. Bad pixels were 3d-Bezier-interpolated in the datacube. Each frame was then wavelength calibrated and converted to a data cube following the ESO standard reduction procedure, based on several `spred`-routines. Differential atmospheric dispersion was corrected by tracing the peak of the stellar PSF through the cubes and correcting for the offsets.

The next important step was the extraction of the spectra from the 79 data cubes on target. The importance of preparing a consistent and comparable data set has already been stressed. Consequently, the largest single extraction aperture centered on the PSF peak was determined to fit all data cubes. Spectra where then extracted slice by slice by integrating over the aperture around the fitted PSF peaks. Due to the rather poor atmospheric conditions, the FWHM of the AO corrected H band PSF was typically 0.5" (5 pixel) while the value for the K-band came out to be 0.2" (2 pixel) about 4 to 10 times larger than the value expected for multicolor diffraction spike. Fig. 5 shows the FWHM of the visual seeing measured by the DIMM seeing monitor in H and K bands during the observations. Fig. 6 displays sample spectra as they were extracted from two of the 79 data cubes.

### 4.2 Aligning the extracted spectra

Residual wavelength jitter between the single spectra of the order of 1/10 of a pixel was corrected by correlating the G0V stellar Brγ lines in every single spectrum. The line is clearly visible in Fig. 6. The residual spectral jitter could thus be lowered to 1/50 – 1/100 of a pixel.

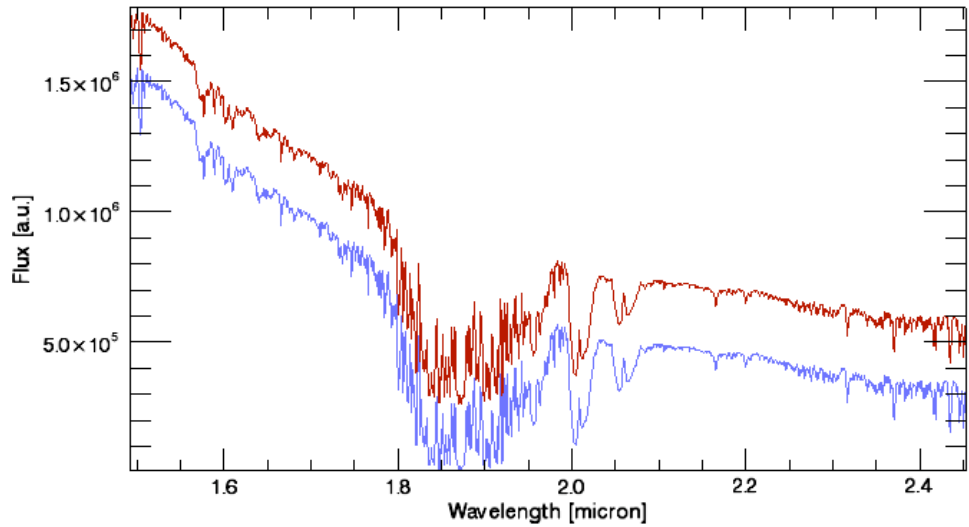

Fig.6. Typical spectra of HD209458 as extracted from two of the 79 data cubes. The upper spectrum has been offset for clarity. The general drop in flux towards longer wavelengths probably reflects the efficiency across the rather broad spectral window. Outside the noisy region between 1.8 and 1.95 μm where the atmospheric transmission vanishes, the noise is significantly smaller than the spectral features, as comparing the two spectra reveals.

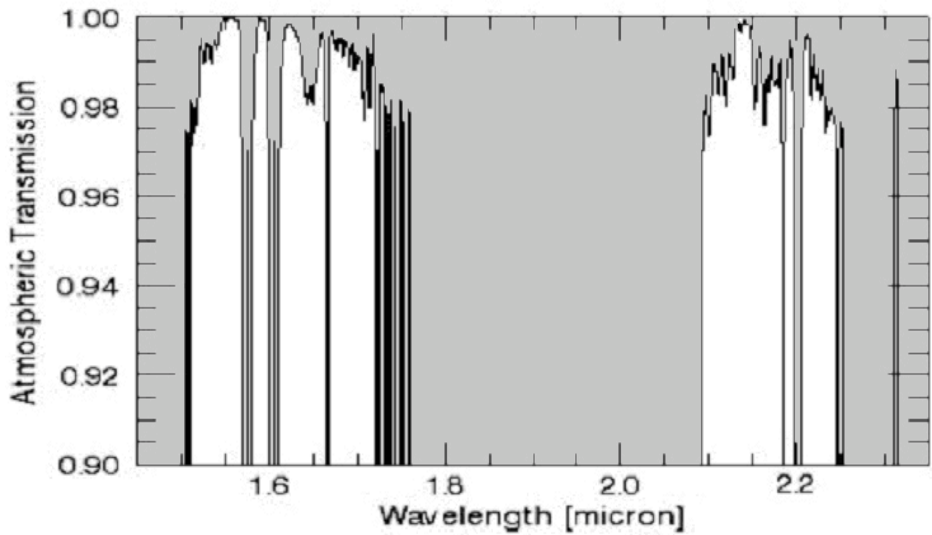

Fig. 7. The atmospheric filter for the correction of the overall transmission was created from a synthetic ATRAN spectrum (Lord 1992) by setting all wavelengths with a transmission lower than 0.97 to zero.

The fluctuation of the overall atmospheric transmission has already been addressed above. Such fluctuations either result from increasing the concentration of water and other chemical species in the atmosphere, which deepen and widen atmospheric absorption bands, or they result from direct obscuration, e.g., scattering by thick clouds. The gross atmospheric transmission can largely be disentangled from the absorption by chemical species by selecting those spectral bands, which are rather insensitive to an increase of the zenith distance. We used a synthetic standard atmospheric spectrum based on the ATRAN code (Lord 1992) and created a spectral box filter indicating those spectral bands with a transmission exceeding 97% shown in Fig. 7. Integrating each stellar spectrum multiplied with such filter provides a reliable indicator of the transmission fluctuations throughout the night. These fluctuations are displayed in Fig. 8 in a ln plot as a function of zenith distance. The total amplitude of transmission variations is about 12% if one also allows for the usual zenith distance dependence of the transmission. Each spectrum was then calibrated with the appropriate factor.

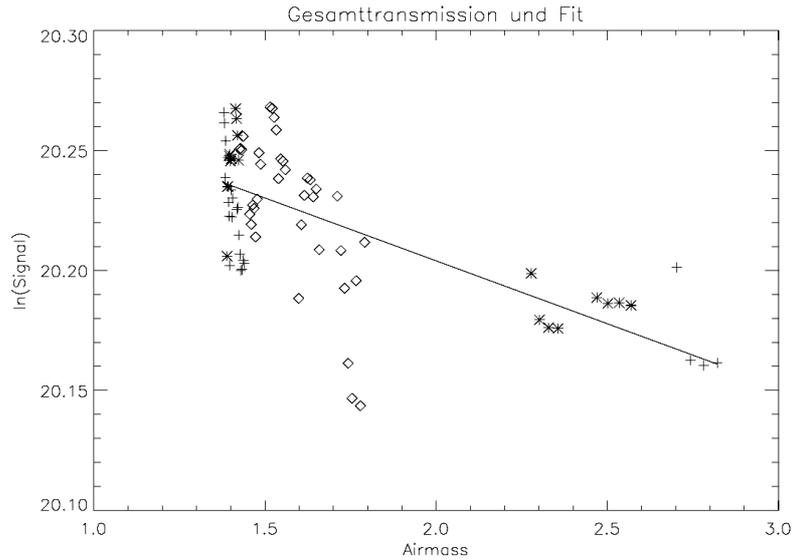

Fig.8. Plot of gross-transmission vs. airmass and an unconstrained linear fit (straight line). Diamonds denote spectra obtained during occultation, + signs those obtained outside occultation and snowflakes those spectra obtained during the transition, i.e., between $1^{st}$ and $2^{nd}$ contact or between $3^{rd}$ and $4^{th}$ contact.

As a next step the spectra were corrected for different air masses. Since each spectral channel has its own absorption characteristics, the individual zenith distance dependences had to be computed and individually corrected for each wavelength channel. The spectra were divided into 2 groups, during occultation and out-of occultation. Both groups were treated separately. A linear unconstrained fit ln(signal) versus airmass fit was used in each spectral channel for both datasets. Typical results are shown in Fig. 9. They demonstrate that the correction generally worked very well. However, residual fluctuations are visible in Fig. 9 (bottom). Very likely they are responsible for the low order fluctuations in the final spectrum (Fig. 10).

After correcting for gross transmission and zenith-distance differences, the spectra of both phases, during occultation and out-of occultation were separately averaged to form two mean spectra. The planet/star contrast was obtained by dividing the out-of occultation spectrum, which includes the planetary signal, by the occultation spectrum. Since the stellar component is contained in both spectra, the ratio will cancel the stellar contribution completely, including stellar light reflected off the planetary atmosphere. It will also account for all instrumental effects. The result should also be corrected for telluric features, but this is true only to a certain degree, given the prevailing conditions. The resulting spectrum did in fact show residual water features, in particular between the H and K band. An differential airmass ATRAN spectrum made up for water only was fitted to the residuals to divide out the residual water lines. We also noted residual OH lines at a level of 0.5% to 1%. The sample points of those spectral intervals affected by OH emission lines were deleted and reset to the values of a polynomial fit through the spectrum. The emerging final H band and K band are displayed in Fig. 10.

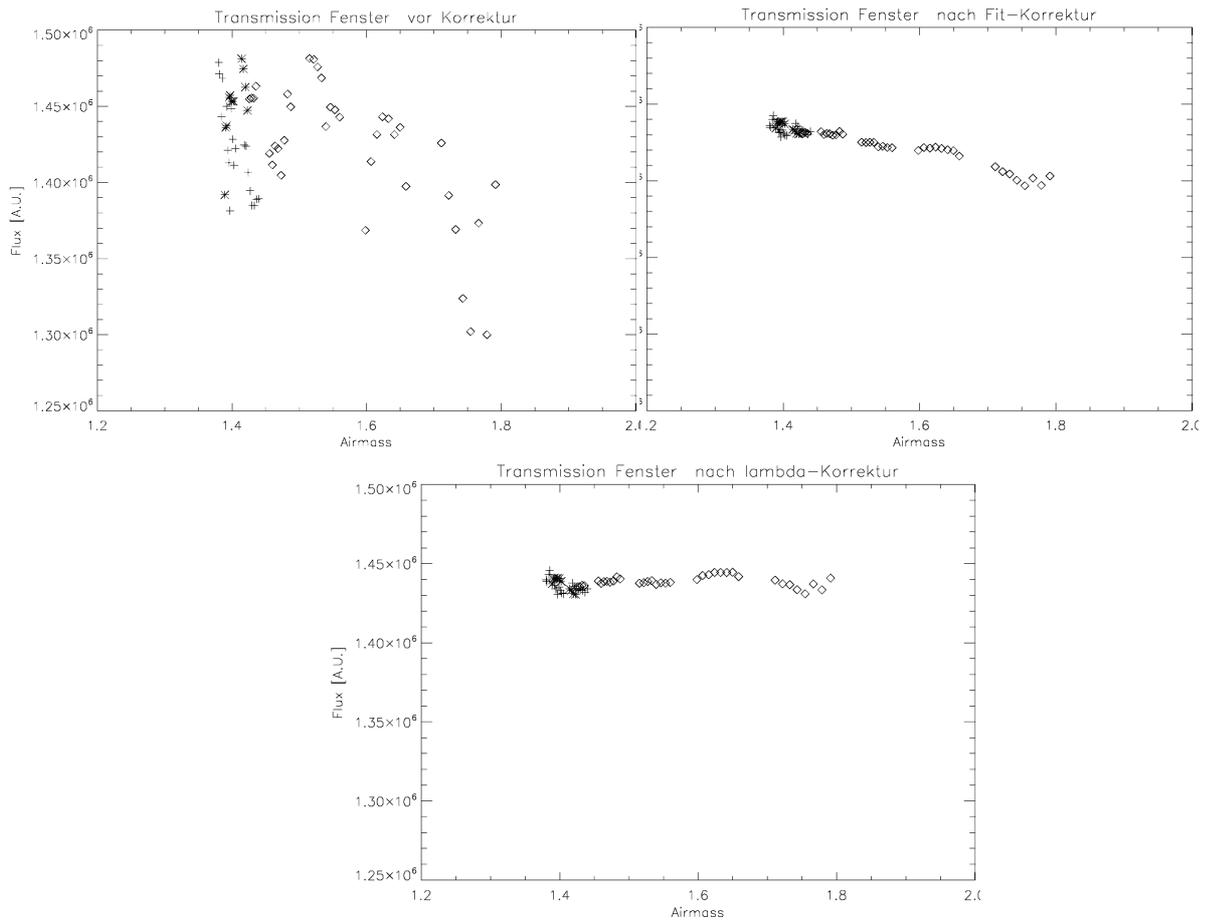

Fig.9. Example of the reduction of a single H-band wavelength channel at 1.5295 μm through all spectra. Upper left: Flux vs. airmass before corrections. Upper right: flux vs. airmass after gross-transmission correction. Bottom: Flux vs. airmass after air mass correction.

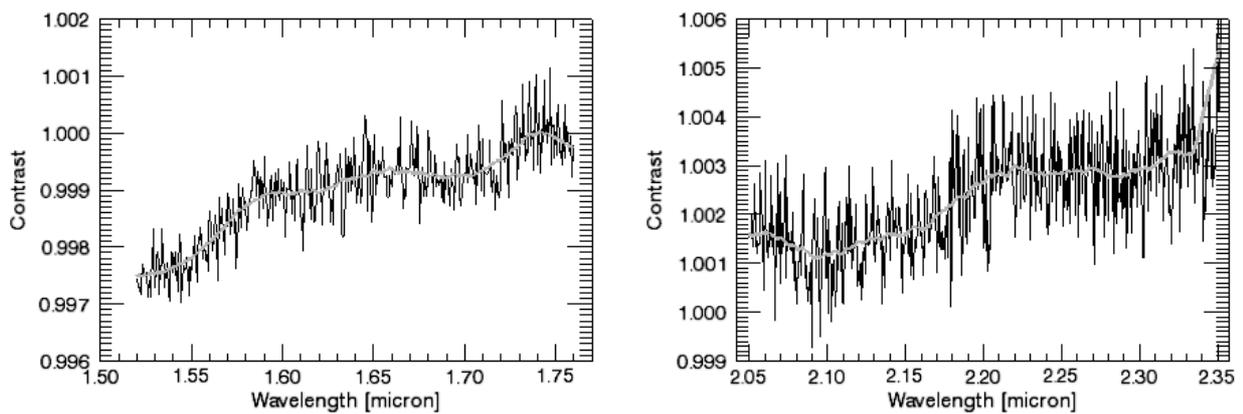

Fig.10. H- and K-Band contrast-spectra, 60-channel smoothed signal (grey). The residual OH lines have been corrected for.

The overall shape of the spectra seems to be dominated by systematic effects, since the baselines of the spectra are varying much more than what can be expected from the exoplanet. The intrinsic noise per sample position, measured relative to a 60-channel smoothed spectrum (Fig. 10), is of the order of 1/2500 in the H-band and 1/1400 in the K band spectrum. These values are very close to the expected numbers if one also accounts for the fact that only a quarter of the expected data was accumulated. We conclude that, in general, single spectral features might not be affected by systematic noise and it might in fact be possible to correlate our data relative to model spectra in the literature. These investigations are under way and will be published in a forthcoming paper (Angerhausen et al. 2006)

## 5. CONCLUSIONS

The feasibility of NIR- imaging spectroscopy as a method to explore extrasolar planetary spectra has been demonstrated to work under even mediocre atmospheric conditions, which also limited our observation time. The final S/N of our data is within the expectation of having obtained only 30 minutes out-of-occultation data (data including containing the planetary signal) if one does not account for gross spectral features. Under more favorite observing conditions and with improved observing efficiency, our goal, isolating an extrasolar planetary spectrum, seems feasible.

## REFERENCES

1. Angerhausen D. et al. 2006, in preparation
2. Arribas et al. 2006, PASP, 118, 21
3. Barman T. et al. 2005, ApJ, 632, 1132
4. Charbonneau D. et al.2005, ApJ, 626, 523
5. Cody, A. M.; Sasselov, D. D. 2002, ApJ, 569, 451
6. Deming D.; Seager S. 2005, Nature, 434, 740
7. Hartung M. et al. 2004 A&A, 421, L17-L20
8. Henry G. W. et al. 2000, ApJ, 529, L41
9. Kissler-Patig M. 2005, Very Large Telescope SINFONI Users Manual
10. Laughlin G. et al. 2005, ApJ, 629, L121
11. Lord S. D., 1992, NASA Technical Memorandum 103957
12. Marcy, G. W.; Butler, R. P. 1995, Bulletin of the AAS, vol. 27, p. 1379
13. Richardson L. J. et al. 2003, ApJ,597, 581
14. Sudarsky D. et al. 2003, ApJ, 588, 1121
15. Wittenmyer R. A. et al. 2005, ApJ, 632, 1157